\documentclass[global,twocolumn]{svjour}
\usepackage{color}
\usepackage{graphics}
\usepackage{amsmath}
\journalname{}
\begin{document}

\title{Simultaneous magneto-optical trapping 
of lithium and ytterbium atoms
towards production of ultracold polar molecules}
\author{M. Okano\inst{1} \and H. Hara\inst{1} \and M. Muramatsu\inst{1} 
\and K. Doi\inst{1} \and S. Uetake\inst{1,2} \and Y. Takasu\inst{1}
\and Y. Takahashi\inst{1,2}
}                     
\offprints{M. Okano, Fax: +81-75-753-3769, E-mail: oquano@scphys.kyoto-u.ac.jp.}
%
\institute{Department of Physics, Graduate School of Science, 
Kyoto University, Kyoto, Japan \and 
CREST, Japan Science and Technology Agency, Kawaguchi, Saitama, Japan}
\date{Received: date / Revised version: date}
\maketitle
\begin{abstract}

We have successfully implemented the first simultaneous 
magneto-optical trapping (MOT) of
lithium ($^6$Li) and ytterbium ($^{174}$Yb) atoms,
towards production of ultracold polar molecules of LiYb. 
For this purpose, we developed the dual atomic oven which contains 
both atomic species as an atom source
and successfully observed the spectra of the Li and Yb atoms in the atomic beams 
from the dual atomic oven.
We constructed the vacuum chamber including the glass cell
with the windows made of zinc selenium (ZnSe) for the CO$_2$ lasers,
which are the useful light sources of optical trapping
for evaporative and sympathetic cooling.
Typical atom numbers and temperatures in the compressed MOT are
7$\times$10$^3$ atoms, 640 $\mu$K for $^6$Li,
7$\times$10$^4$ atoms and 60 $\mu$K for $^{174}$Yb, respectively. 

\end{abstract}

\section{Introduction}
\label{intro}

Researches of ultracold gases of atoms have revealed 
the nature of the quantum degenerate gases
since the first realization
of the Bose-Einstein condensates (BEC) \cite{{Rb},{Na}}
and the Fermi degenerates (FD) \cite{FDG} of atoms.
As a next important step, 
an ultracold gas of molecules has recently attracted much attention 
due to their unique properties and prospects for many applications \cite{Doyle}. 
In particular, molecules consist of two atomic species with different mass
may have large electric dipole moments, called polar molecules. 
Electric dipole-dipole interactions of polar molecules
are anisotropic and in a long-range,
different from the contact interactions of atoms.
In addition, dipole-dipole interactions of electric dipole moments
are considered to be much stronger than those of magnetic dipole moments of atoms. 
Due to these strong long-range anisotropic interactions, 
many applications of polar molecules have been theoretically proposed \cite{book}. 
Polar molecules in optical lattices have particularly been 
worthy of attention for their applications including
a quantum computation \cite{DeMille}, 
the study of the frustrated states in the triangular lattices \cite{Lewenstein_nature}
and the quantum simulator of lattice-spin models \cite{Zoller}. 

Among the various combinations of atomic species, 
we have focused on ultracold gases of molecules 
composed of Li and Yb \cite{Okano}. 
The molecule has the largest mass ratio ($\sim$29) 
among the combinations of atomic species 
that have already been cooled to quantum degenerate regime. 
The most important advantage of molecules of LiYb is 
the existence of a spin degree of freedom in the electronic ground molecular state
($^2\Sigma$), 
which enables us to implement the quantum simulator of lattice-spin models.
In contrast, molecules composed of alkali atoms
in the singlet electronic ground molecular state ($^1\Sigma$)
have no electron spin degree of freedom. 
RbYb molecules, as well as LiYb molecules, 
have a spin degree of freedom in the $^2\Sigma$ state
and have recently been studied \cite{RbYb}.
However, LiYb molecules have further advantages:
(i) 
Li and Yb have both bosonic and fermionic isotopes
and this leads to the study of various molecules
composed of fermionic-fermionic, fermionic-bosonic and bosonic-bosonic atoms.
In contrast, Rb has only bosonic isotopes.
(ii)
Due to their large mass ratio,
weakly bound LiYb molecules of both fermionic isotopes 
are expected to be collisionally stable
and promising for the study of the three-body system
such as Efimov trimer states \cite{Petrov}.

The expected processes for generating ultracold molecules of LiYb are as follows: 
First, Li and Yb atomic beams from an atomic oven 
are decelerated by the Zeeman slowing method and 
these cooled atoms are loaded to a magneto-optical trap (MOT)
in a main chamber. 
Second, these trapped atoms are transferred to a optical trap 
and evaporatively cooled with sympathetic cooling of atoms
between different atomic species \cite{Sympathetic}. 
Third, evaporatively cooled Li and Yb atoms
are adiabatically associated to weakly bound LiYb molecules
by employing magnetically tunable Feshbach resonances \cite{Feshbach}. 
Finally, weakly bound molecules are coherently transferred to 
the rovibrational ground state of the electronic ground molecular state 
by employing a step of STIRAP (STImulated Raman Adiabatic Passage) \cite{KRb}.

In this paper, 
we report the first simultaneous MOT of Li and Yb atoms,
towards production of ultracold polar molecules of LiYb. 
For this purpose, 
we developed the dual atomic oven which contains both atomic species
and successfully observed the spectra of Li and Yb atomic beams 
from the dual atomic oven as described in Sect. \ref{dual atomic oven}.
The laser systems and a glass cell as a main chamber
are described in Sect. \ref{experimental set up}.
In particular,
the windows made of zinc selenium (ZnSe) are attached to the glass cell
for the CO$_2$ laser,
which is a useful light source of optical trapping
for evaporative and sympathetic cooling
as well as trapping of cold LiYb molecules.
The procedures and experimental parameters 
of the simultaneous MOT of $^6$Li and $^{174}$Yb 
are discussed in Sect. \ref{experiment}.

\section{Dual atomic oven}
\label{dual atomic oven}

\begin{figure*}[ht]
\begin{center}
\resizebox{0.85\textwidth}{!}{%
  \includegraphics{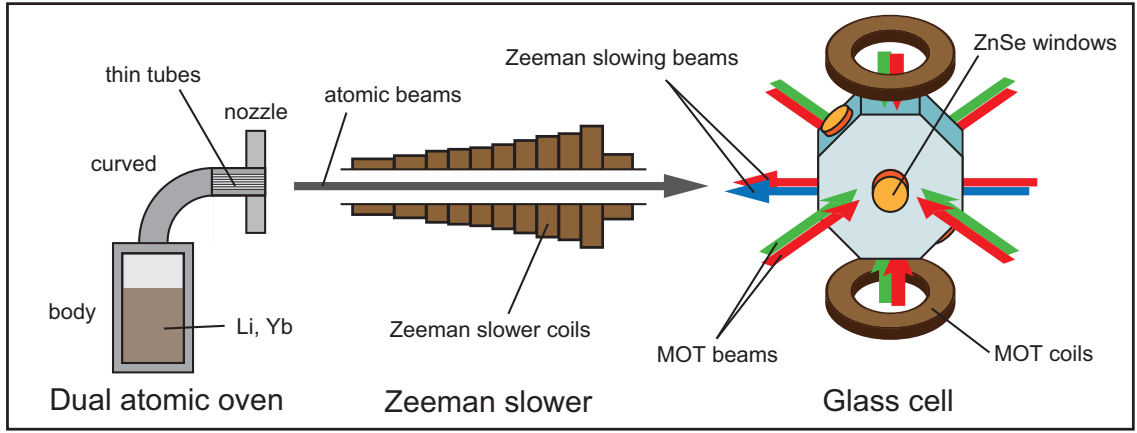}
}
\caption{Schematic setup of the experiment.
The Li and Yb metals are put in the dual atomic oven.
Li and Yb atomic beams from the dual atomic oven 
are decelerated through the Zeeman slower coils
by applying Zeeman slowing beams.
These atoms are loaded to a MOT
in the glass cell with ZnSe windows.
MOT beams propagate along the three axes 
and the MOT coils are set in the vertical axis.
}
\label{fig:set_up}       
\end{center}
\end{figure*}

\subsection{Atomic oven}
\label{atomic oven}

We first developed the dual atomic oven which contains both Li and Yb atoms
as an atom source.
Developing the dual atomic oven is important 
to make experimental systems simple and compact.
While temperatures of Li and Yb atoms are equal in the dual atomic oven, 
the saturated vapor pressure of Yb is 
at most an order of magnitude larger than 
that of Li at around 400 $^{\circ}$C.
The configuration of the dual atomic oven 
is shown in the left side of Fig. \ref{fig:set_up}.
This oven consists of three parts: the body part which contains Li and Yb atoms,
the curved part and the nozzle part 
where atoms pass out through as the atomic beams.
We designed the body part vertical 
because it is heated above the melting point of Li (180 $^{\circ}$C).
Approximately 60 thin tubes, each of which has a inner diameter of 0.3 mm and 
a length of 10 mm, are assembled at the nozzle part, 
which limits the divergence of the atomic beams 
by approximately 0.03 rad.

\subsection{Spectroscopy using the atomic beams}
\label{spectroscopy}

One may expect that atoms could not pass out as the atomic beams
because of reactions between Li and Yb in the atomic oven.
To check the performance of the dual atomic oven, 
we employed the spectroscopy of Li and Yb atomic beams from the dual atomic oven.
The light source for the Li spectroscopy is
an extended cavity laser diode (ECLD) in the Littman configuration  
with a wavelength of 671 nm.
For the Yb spectroscopy we also used an ECLD in the Littrow configuration 
with a wavelength of 399 nm.
The Li and Yb metals with natural abundances 
were put in the body part of the atomic oven,
which was heated to 460 $^{\circ}$C by wires wound around.
The vapor pressures of Li and Yb at this temperature are 
$1\times10^{-3}$ Torr and $7\times10^{-3}$ Torr, respectively.
The output laser beams of ECLDs with their laser frequencies tuned to
$2s$-$2p$ transitions of Li and 
($6s^2$) $^1S_0$-($6s6p$) $^1P_1$ transitions of Yb
were applied perpendicular to the atomic beams and
fluorescences from the atomic beams
were detected by a photo multiplier tube (PMT). 

\begin{figure}
\begin{center}
\textbf{a}
\resizebox{0.45\textwidth}{!}{%
  \includegraphics{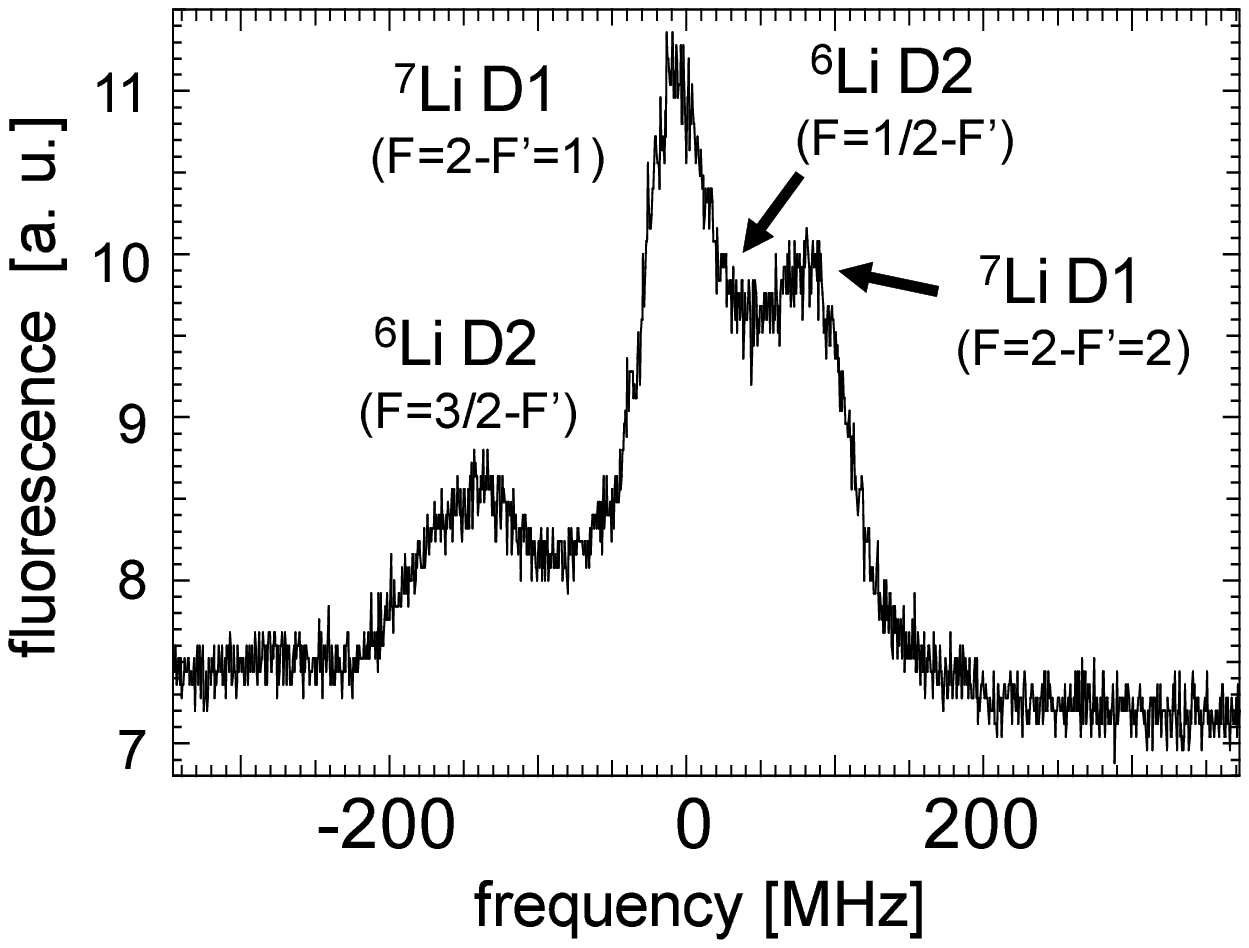}
}\\
\textbf{b}
\resizebox{0.45\textwidth}{!}{%
  \includegraphics{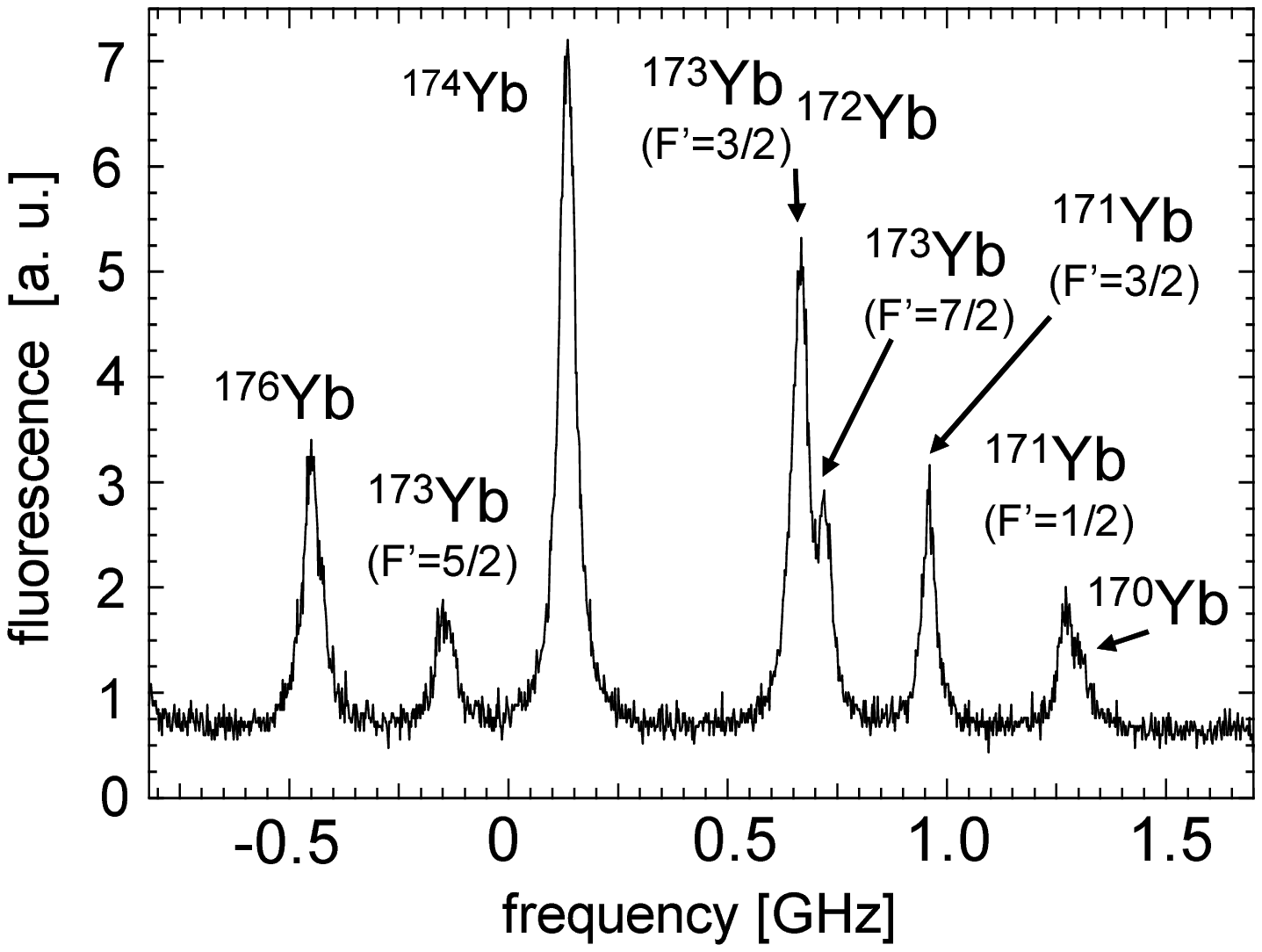}
}
 
\caption{The fluorescence signals of 
the atomic beam of Li(\textbf{a}) and Yb(\textbf{b}).
The Li and Yb metals with natural abundances were used in this measurement.
Transitions of the $D_2$ line of $^6$Li and the $D_1$ line of $^7$Li
($\lambda$ = 671 nm) are observed (16 times integration)(\textbf{a}) and
($6s^2$) $^1S_0$-($6s6p$) $^1P_1$ transitions of Yb isotopes 
($\lambda$ = 399 nm) are observed(\textbf{b}).
Each peak of signals is indicated 
by an isotope and a transition and
$F$ and $F^{\prime}$ denote hyperfine states of the ground state ($S$) and
the excited state ($P$), respectively.
}
\label{fig:Li_signal} 
\end{center}
\end{figure}

In this way we observed the spectra of Li and Yb atomic beams 
by measuring the fluorescence as shown in Fig. \ref{fig:Li_signal}.
The fluorescence signals of the $D_2$ line of $^6$Li and the $D_1$ line of $^7$Li
in Fig. \ref{fig:Li_signal}(a)
were obtained with 16 times integration
and those of $^1S_0$-$^1P_1$ transitions of Yb isotopes were 
also obtained (Fig. \ref{fig:Li_signal}(b)).
The Doppler widths of the fluorescence signals 
resulted from the angular divergence of the atomic beams
are consistent with the configuration of the thin tubes 
at the nozzle part of the dual atomic oven.
Thus the performance of the dual atomic oven was properly checked and 
the dual atomic oven was prepared for the experiment.
The relative intensities of the each lines are consistent with the expectation
\cite{Li_spec}.

\section{Experimental setup}
\label{experimental set up}

\subsection{Laser systems}
\label{laesr systems}

\begin{figure}
\begin{center}
\textbf{a}
\resizebox{0.4\textwidth}{!}{%
  \includegraphics{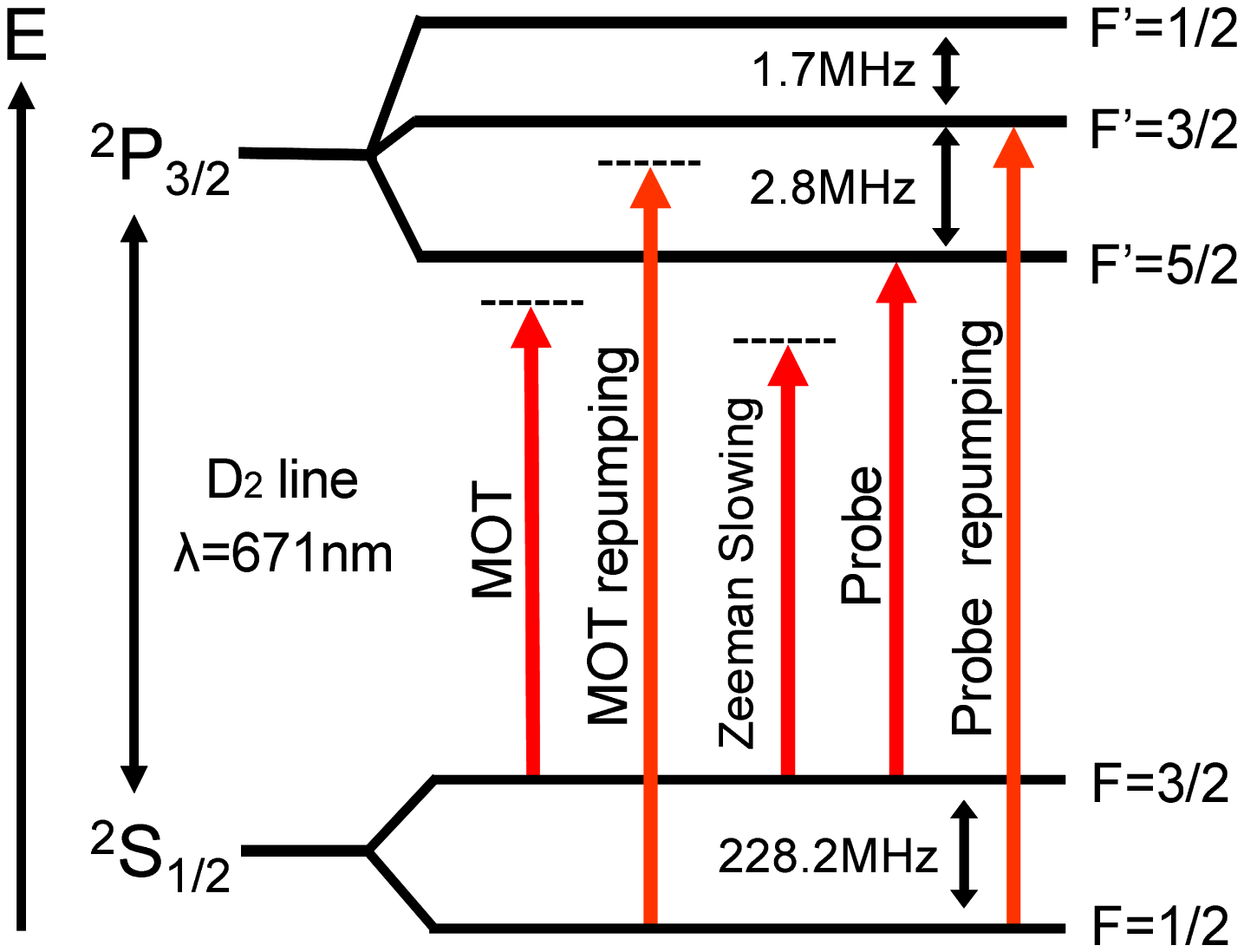}
}\\
\textbf{b}
\resizebox{0.4\textwidth}{!}{%
  \includegraphics{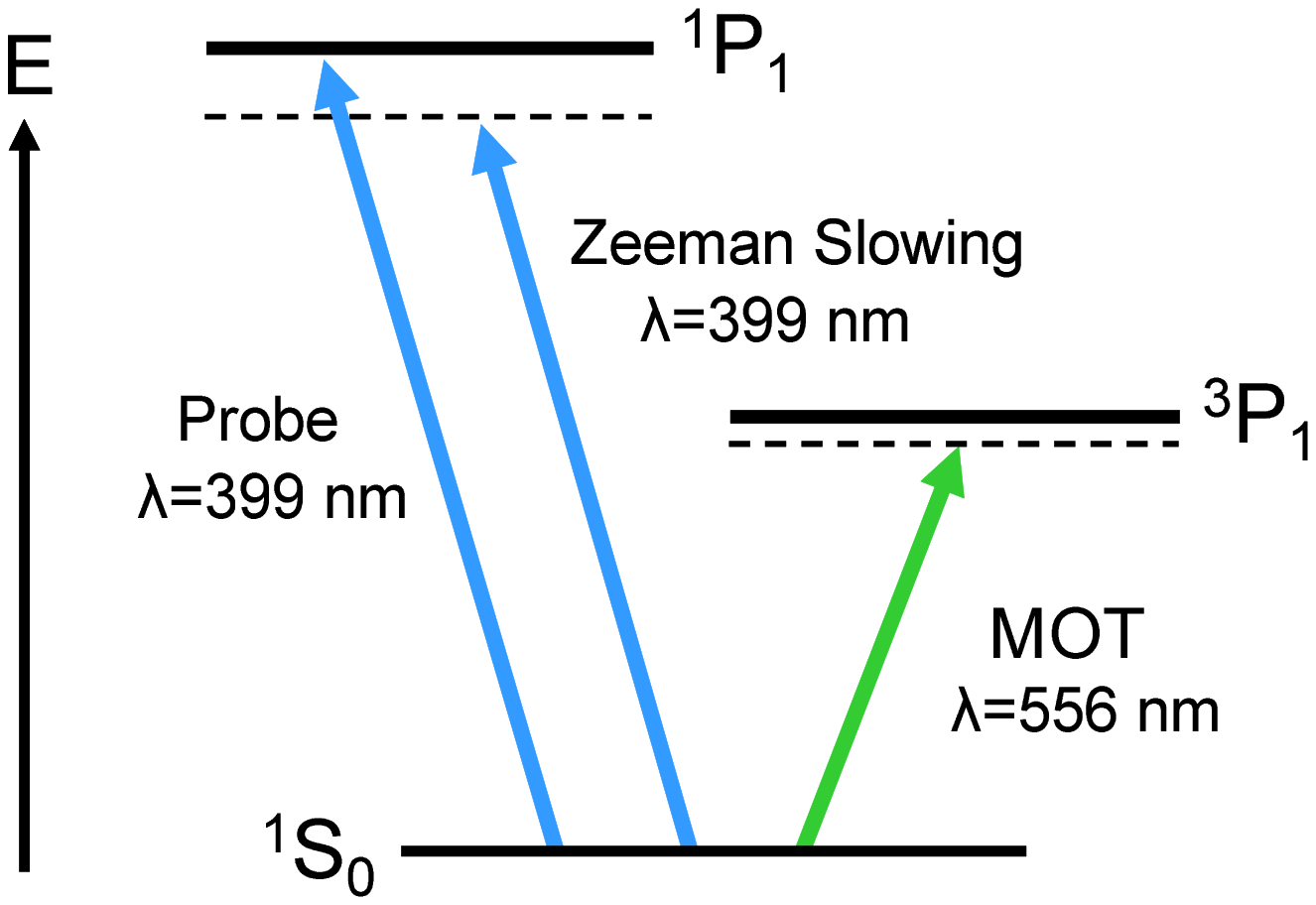}
}
\caption{Energy levels and transitions for laser cooling and probing
of $^6$Li (\textbf{a}) and $^{174}$Yb (\textbf{b}) atoms.
}
\label{fig:transitions} 
\end{center}
\end{figure}

Energy levels and transitions for laser cooling and probing 
of $^6$Li and $^{174}$Yb atoms are shown in Fig. \ref{fig:transitions}.
We use transitions of the $D_2$ line of $^6$Li atoms
for laser cooling and probing.
The wavelength and the natural linewidth ($\Gamma$) of the $D_2$ line
are 671 nm and 2$\pi\times$5.9 MHz, respectively. 
As a light source we used a cw ring dye laser (Coherent, 899-21) with LD688 dye
pumped by a cw green diode-pumped solid-state laser (Coherent, Verdi-V8)
with a wavelength of 532 nm and a power of 6 W.
The output power of the cw ring dye laser is 400 mW. 
The output beam of the dye laser is divided into laser beams
for MOT, MOT repumping, Zeeman slowing, probe, probe repumping, 
and frequency locking of the dye laser.
The use of repumping beams (MOT repumping beam and probe repumping beam)
is essentially important
because the hyperfine splittings of the excited state ($^2P_{3/2}$) 
are comparable with the natural linewidth.
The frequency of each laser beam is shifted by using
acousto-optic modulators (AOM).
The frequency of the dye laser is locked to the transfer cavity, 
which is a 50cm-length confocal cavity whose spacer is made of invar,
a nickel steel alloy with low thermal expansion.
Furthermore the transfer cavity is locked to 
the 556 nm laser light whose frequency is highly stabilized 
for laser cooling of Yb atoms as described below.  
The powers of MOT beam and MOT repumping beam are 11 mW and 12 mW, 
respectively and each of these beams is divided into three beams
with a diameter of 10 mm, which produce
the average peak intensities of each MOT beam and MOT repumping beam
of 1.4 $I_s$ and 1.6 $I_s$, respectively
where $I_s$ is the saturation intensity of the $D_2$ line, 2.54 mW/cm$^2$.  
The power of Zeeman slowing beam is 60 mW.
MOT repumping beam is also used as probe repumping beam and
the frequencies of the probe beam and probe repumping beam
are set to the resonances of the transitions.

We use the transitions of $^1S_0$-$^1P_1$ and $^1S_0$-$^3P_1$ 
for laser cooling and probing of $^{174}$Yb atoms.
Zeeman slowing of the atomic beam is done
by using the strongly allowed $^1S_0$-$^1P_1$ singlet transition, 
whose wavelength and natural linewidth are 
399 nm and 2$\pi \times$ 29 MHz, respectively.
The LD chip is installed to the ECLD in the Littrow configuration
and the wavelength is tuned to 798 nm by grating feedback.
The output beam of the ECLD with a power of 10 mW is
amplified to 480 mW by the TA with the operating current of 1.5 A. 
To convert the amplified laser light to 399 nm light,
we used a periodically poled potassium titanyl phosphate (PPKTP) nonlinear crystal.
The PPKTP crystal is put in a crystal oven made of copper 
whose temperature is stabilized by the Peltier unit.
This crystal oven is set in a ring cavity 
to enhance the efficiency of second harmonic generation (SHG) and
399nm laser beam with a power of 40 mW is generated.
The frequency of the 798 nm ECLD is locked to the transfer cavity
and shifted by using the AOM.

Probing of $^{174}$Yb atoms is also done
by using the $^1S_0$-$^1P_1$ transition. 
Because we need a frequency stable laser light with
good transverse mode to properly probe atoms,
we adopted an ECLD with optical feedback to a filter cavity \cite{Hayasaka}.
The ECLD is constructed in the Littrow configuration and
the wavelength and the standard output power of the LD chip (NICHIA) 
are 399nm and 65 mW, respectively.
The output beam of the ECLD with a power of 4.5 mW is injected to the filter cavity
and then partially retro-reflected ($\approx$ 4\verb|%|) 
by a wedge plate to stabilize the laser frequency by optical feedback.
In addition long-term stability of the frequency of the laser is obtained
by 1f-3f electric feedback with modulation frequency ($f$) of 40 kHz.
The output beam in $\textrm{TEM}_{00}$ mode is obtained with a power of 2.1 mW .
To lock the frequency of this laser system,
we constructed a vacuum chamber with an atomic oven of Yb.
The frequency stabilized output beam of the filter cavity is partially divided and
applied to the atomic beam from the atomic oven, 
which contains 50 g of the natural Yb metals.
The fluorescences from the atomic beam of Yb are detected by the PMT and
the frequency of the laser system is locked to the resonance of $^{174}$Yb.

MOT of $^{174}$Yb atoms is done by using the intercombination $^1S_0$-$^3P_1$ 
transition \cite{kuwamoto}, whose wavelength and natural linewidth are
556 nm and 2$\pi \times$ 182 kHz, respectively.
The natural linewidth is so narrow that 
we use the high power narrow linewidth laser with a wavelength of 556 nm \cite{556}
as a light source.
This laser system consists of a commercial 1 W fiber laser 
and a lithium triborate (LBO) nonlinear crystal in a ring cavity for SHG.
The ytterbium-doped fiber laser with a wavelength of 1112 nm 
is frequency locked to a ultra-low-expansion (ULE) cavity 
mounted in a temperature-controlled vacuum chamber.
This laser system generates 556 nm laser light 
with a linewidth of less than 100 kHz.
The transfer cavity is locked to 
this frequency stabilized 556 nm laser light as described above.
The power of MOT beam is 75 mW and divided into three beams
with a diameter of 16 mm.
This means that the average peak intensity of each MOT beam is about 90 $I_s$
where $I_s$ is the saturated intensity 
of the $^1S_0$-$^3P_1$ transition, 0.14 mW/cm$^2$.

\subsection{Glass cell}
\label{glass cell}

A main experimental region needs some conditions.
It needs good optical accessibility for many laser beams 
used for laser cooling and probing of both atoms.
It also needs to be compact 
because MOT and Feshbach coils put closer to the center of the main chamber 
can yield larger magnetic fields.
To meet these conditions, 
we designed a octagonal glass cell made of TEMPAX Float (SCHOTT) as the main chamber
as shown in the right side of Fig. \ref{fig:set_up}.
The pressure in the glass cell is estimated to be
an order of 10$^{-9}$ Torr.

As light sources for optical trapping we chose the CO$_2$ lasers 
with a wavelength of 10.6 $\mu$m.
The CO$_2$ lasers have significantly longer wavelength than atomic transitions of 
Li and Yb, therefore electric fields of the CO$_2$ laser beams
are considered to be quasi-static electric field. 
This means that the CO$_2$ lasers are useful for trapping Li and Yb atoms 
as well as LiYb molecules. 
For the CO$_2$ laser beams, we attached four windows made of ZnSe that are AR coated
at 10.6 $\mu$m by using the adhesion bond (Epoxy technology, EPO-TEK 353ND) 
compatible with the ultra high vacuum.

\section{Simultaneous magneto-optical trapping 
of Li and Yb atoms}
\label{experiment}

In the experiment, more than 95\verb|%|
$^6$Li enriched Li metals and the natural Yb metals are put
in the dual atomic oven, whose body part is heated to 400 $^{\circ}$C.
Li and Yb atomic beams from the dual atomic oven 
are decelerated through the Zeeman slower coils
by applying Zeeman slower beams of Li and Yb atoms as shown 
in the middle of Fig. \ref{fig:set_up}.
The Zeeman slower coils are in a increasing magnetic field configuration 
along the atomic beams 
and the maximum value of the magnetic field is designed approximately 520 G.
Both Zeeman slowing beams are gradually focused 
towards the nozzle part of the dual atomic oven
in accordance with the angular divergence of the atomic beams. 
The Zeeman slowed Li and Yb atoms are 
then loaded to a MOT in the glass cell.
MOT beams propagate along the three axes 
and the MOT coils (anti-Helmholtz coils) are set in the vertical axis
as shown in the right side of Fig. \ref{fig:set_up}.
The axial (radial) magnetic field gradients for 
the MOT are 16 G/cm (8 G/cm).
The detunings of MOT beam, MOT repumping beam of Li and MOT beams of Yb
are 24 MHz, 24 MHz ($\sim$ 4.1 $\Gamma$) 
and 4 MHz ($\sim$ 22 $\Gamma$), respectively.  
Zeeman slowing and MOT beams of Li and Yb atoms are 
overlapped at dichroic mirrors 
and then go through quarter wavelength plates 
designed for both wavelengths of laser beams in front of the glass cell. 
In these conditions, 
we have successfully observed the fluorescences from the simultaneous MOT of 
$^6$Li and $^{174}$Yb atoms by a charge-coupled-device (CCD) camera.
Loading of atoms to the simultaneous MOT is saturated within a few seconds,
which is mainly limited by collisions of atoms with the background gases.

\begin{figure}
\begin{center}
\textbf{a}
\resizebox{0.20\textwidth}{!}{%
  \includegraphics{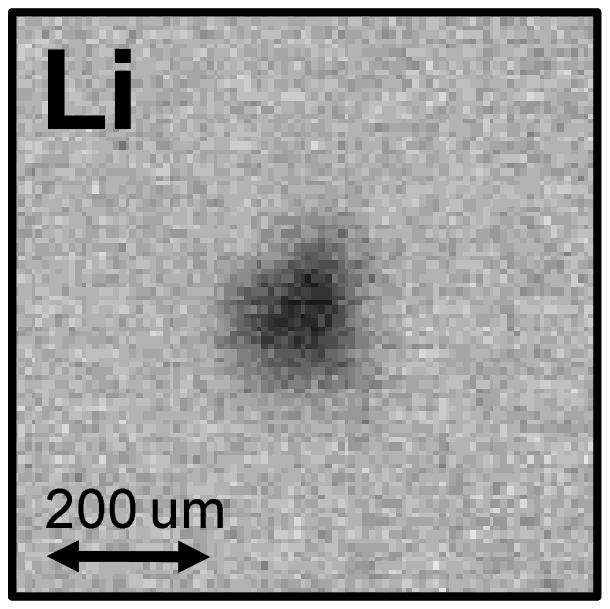}
}
\textbf{b}
\resizebox{0.20\textwidth}{!}{%
  \includegraphics{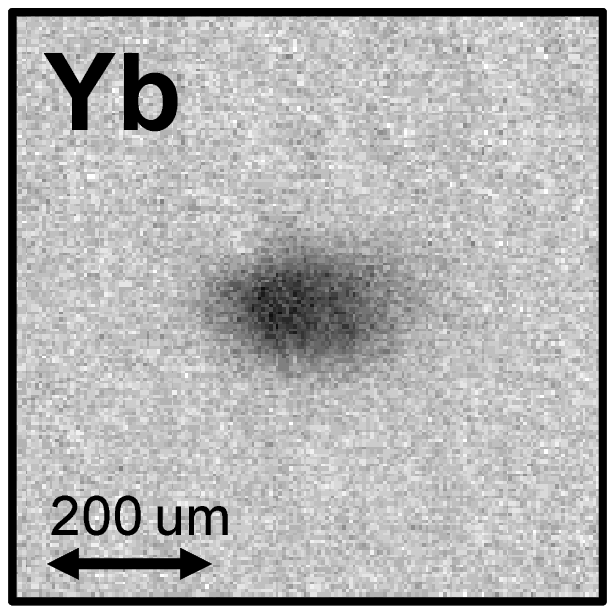}
}
\caption{Absorption images of $^6$Li atoms(\textbf{a}) and 
$^{174}$Yb atoms(\textbf{b}) in the compressed MOT.
These density profiles of atoms right after turning off the MOT beams
were obtained by absorption imaging technique}
\label{fig:TOFs} 
\end{center}
\end{figure}

To increase the density and decrease the temperature of the atoms,
we have implemented the compressed MOT for both atoms 
by increasing the magnetic field gradients
and decreasing the detunings and the intensities of MOT beams.
After the 2.5 s loading of atoms to the MOT, 
Zeeman slowing beams are shut off and
the detunings of MOT beams, MOT repumping beams of Li and MOT beams of Yb
are switched to 12 MHz, 12 MHz ($\sim$ 2.0 $\Gamma$) 
and 1 MHz ($\sim$ 5.5 $\Gamma$), respectively.
Then the axial (radial) magnetic field gradients are linearly increased 
to 30 G/cm (15 G/cm) during a sweeting time of some tens of ms.
After ramping the magnetic field gradients, 
the intensities of MOT beams of Li and Yb
are decreased to approximately 1/10 and 1/20, respectively
and the magnetic field gradients are kept at the maximum value 
during the cooling time.
The atom number and the temperature of atoms in the compressed MOT are measured
by the time-of-flight (TOF) method.
The powers of probe beams of $^6$Li and $^{174}$Yb are
35 $\mu$W and 50 $\mu$W, respectively and
the diameters of both probe beams are 6 mm.
The density profiles of atoms right after turning off the MOT beams
were obtained by absorption imaging technique as shown in Fig. \ref{fig:TOFs}. 
Typical values of the compressed MOT are as follows :
The atom number ($N$) and the temperature ($T$) of the $^6$Li 
in the compressed MOT
are $N=7\times10^3$ and $T=640$ $\mu$K.
Those values for $^{174}$Yb
are $N=7\times10^4$ and $T=60$ $\mu$K.
These experimental parameters are listed in Table. \ref{tab:1}.
Smaller number of Li atoms in the trap compared with Yb atoms 
is mainly originated from the fact that 
the configuration of the Zeeman slower was optimized for Yb atoms 
because the larger number of Yb atoms is necessary 
to sympathetically cool Li atoms through the collision 
between Li and Yb atoms in the optical trap.
So far, we could not observe any additional disadvantage
of simultaneous trapping 
on each atom number of Li and Yb in the MOT.

\begin{table}
\begin{center}
\caption{Experimental parameters of the MOT of $^6$Li and $^{174}$Yb:
Wavelength($\lambda$) in vacuum, natural line width($\Gamma$), 
saturation intensity($I_s$), 
detunings($\Delta$) of MOT beam (and MOT repumping beam of Li)
and typical values of the atom number($N$) and the temperature($T$) 
of $^6$Li and $^{174}$Yb atoms in the compressed MOT.
}
\label{tab:1}       
\begin{tabular}{lccc}
\hline\noalign{\smallskip}
& \multicolumn{2}{c}{$^6$Li} & $^{174}$Yb   \\
  & MOT & repumping &  MOT  \\
\noalign{\smallskip} \hline \hline \noalign{\smallskip}
$\lambda$ (nm) & \multicolumn{2}{c}{670.977} &  555.8 \\
$\Gamma$/2$\pi$ (MHz) & \multicolumn{2}{c}{5.9} & 0.182\\
$I_s$ (mW/cm$^2$) & \multicolumn{2}{c}{2.54} &  0.14 \\
$I/I_s$ & 1.4 &1.6 & 90 \\
$\Delta/\Gamma$ & 4.1 & 4.1 & 22 \\
$N$ & \multicolumn{2}{c}{7 $\times$10$^3$} & 7 $\times$10$^4$ \\
$T$ ($\mu$K) & \multicolumn{2}{c}{640}  & 60 \\
\noalign{\smallskip}\hline
\end{tabular}
\end{center}
\end{table}

\section{Conclusion}
\label{conclusion}

In summary, 
we have implemented the first simultaneous MOT of $^6$Li and $^{174}$Yb atoms. 
The dual atomic oven which contains both atomic species was developed
and we have successfully observed the spectra of Li and Yb in the atomic beams 
from the dual atomic oven.
Typical values of atom numbers and temperatures of the compressed MOT are
$7\times10^3$ atoms, 640 $\mu$K for $^6$Li,
$7\times10^4$ atoms and 60 $\mu$K for $^{174}$Yb, respectively. 
The realization of the simultaneous MOT is the key
to production of ultracold polar molecules of LiYb.

The improvement of the vacuum of the main chamber
will reduce the atom losses of the MOT due to the background gases 
and this will increase the atom number of the simultaneous MOT.
In addition to this improvement, 
further optimization of the compressed MOT
are expected to decrease temperatures of the compressed MOT
closer to the Doppler cooling limits of $^6$Li and $^{174}$Yb,
$T=140$ $\mu$K and $T=4$ $\mu$K, respectively.
These improvements will increase the phase space density of the simultaneous MOT 
and this leads to the next process, optical trapping by CO$_2$ lasers.
We can examine the collisional properties of Li and Yb in the optical trap
and obtain the information for production of LiYb molecules.

\begin{acknowledgement}

We thank H. Sadeghpour, P. Zhang, G. Gopakumar, and M. Abe
for helpful discussions
and acknowledge K. Hayasaka, H. Ohira, K. Uchida, and S. Yabunaka
for experimental assistance.
This work was partially supported 
by Grant-in-Aid for Scientific Research of JSPS (No. 18204035) 
and the Global COE program "The Next Generation of Physics, 
Spun from Universality and Emergence" 
from the Ministry of Education, Culture, Sports, Science and Technology
(MEXT) of Japan.
M. O. acknowledges support from JSPS.

\end{acknowledgement}

%

\end{document}